\documentclass[review]{elsarticle}

\usepackage{graphicx,calc,color,subfigmat,amssymb,amsmath,dsfont,epstopdf,tikz}
\usepackage{multirow}
\usepackage{lipsum}
\usepackage{subfigure}
\usepackage{graphicx}
\usepackage{amsmath,amssymb,amsfonts,bm}
\usepackage{breqn}
\usepackage{array}
\usepackage{algorithm}
\usepackage[noend]{algpseudocode}
\usepackage{float}

\usepackage{color}

\journal{Journal of \LaTeX\ Templates}

\begin{document}

\begin{frontmatter}

\title{An effective clustering method based on data indeterminacy in neutrosophic set domain\tnoteref{mytitlenote}}

\author[mymainaddress]{Elyas Rashno }


\author[mymainaddress]{Behrouz Minaei-Bidgoli}

\author[mysecondaryaddress]{Yanhui Guo \corref{mycorrespondingauthor}}
\cortext[mycorrespondingauthor]{Corresponding author.  Department of Computer Engineering, Iran University of Science and Technology, Narmak, Tehran 1684613114, Iran, Email: elyas.rashno@gmail.com; Tel.: +98-916-363-1036.
 }


\address[mymainaddress]{Iran University of Science and Technology, Department of Computer Engineering, Tehran, Iran.}
\address[mysecondaryaddress]{University of Illinois at Springfield, Department of Computer Science, Springfield, Illinois, USA.}

\begin{abstract}
In this work, a new clustering algorithm is proposed based on neutrosophic set (NS) theory. The main contribution is to use NS to handle boundary and outlier points as challenging points of clustering methods. In the first step, a new definition of data indeterminacy (indeterminacy set) is proposed in NS domain based on density properties of data.  Lower indeterminacy is assigned to data points in dense regions and vice versa. In the second step, indeterminacy set is presented for a proposed cost function in NS domain by considering a set of main clusters and a noisy cluster. In the proposed cost function, two conditions based on distance from cluster centers and value of indeterminacy, are considered for each data point. In the third step, the proposed cost function is minimized by gradient descend methods. Data points are clustered based on their membership degrees. Outlier points are assigned to noise cluster; and boundary points are assigned to main clusters with almost same membership degrees. To show the effectiveness of the proposed method, three types of datasets including diamond, UCI and image datasets are used. Results demonstrate that the proposed cost function handles boundary and outlier points with more accurate membership degrees and outperforms existing state of the art clustering methods in all datasets.
\end{abstract}

\begin{keyword}
\texttt{Data clustering, neutrosophic theory, data indeterminacy, image segmentation.}
\end{keyword}

\end{frontmatter}


\section{Introduction}
Clustering is a division of data into groups. Each group, referred as a cluster, attempts to satisfy two rules: objects are similar (or related) to each other (minimize the intra-cluster distance) inside same groups and at the same time different from (or unrelated to) the other groups (maximize the inter-cluster distance) . Clustering represents and models data by few clusters which achieves simplification to data analysis. Data clustering is an important field in machine learning, and has found numerous applications in computer vision, image processing, taxonomy, medicine, geology, business, and pattern recognition community \cite{ClusteringSurveyxu2005survey,ESWA1_khan2013cluster,ESW2Aheloulou2017automatic,ESWA3saha2016brain,ESWA4ramon2017cluster}.

\par In data analysis, clustering methods can be considered as two popular categories: hard (crisp) and fuzzy methods \cite{6-7_baraldi1999survey}. In hard clustering methods, data points are grouped so that each one belongs to exactly one cluster. Unlike hard clustering, in fuzzy partitioning, each object may be assigned to all clusters with different degrees of membership \cite{19_guo2015ncm}. One of the most popular hard clustering methods is K-means which partitions the data into k clusters automatically attempting to minimize the within-group sum of square distances \cite{23_webb2003statistical}. The main disadvantage of this algorithm is that it does not ensure a global minimum of variance and needs a predefined cluster numbers \cite{24_wagstaff2001constrained}.  K-means++ is an improvement of clustering analysis on k-means \cite{25_arthur2007k}. It improves the k-means with choosing the initial cluster values (or "seeds"). It was proposed as an approximation algorithm for the NP-hard k-means problem. K-medoids is also a variation of k-means where it calculates the median for each cluster to determine its centroid. It has the effect of minimizing error over all clusters with respect to the 1-norm distance metric, which relates directly to the k-means algorithm. The Euclidean distance between points is considered as a criterion for clustering and a point designated as the center of that cluster.

\par Fuzzy c-means (FCM) clustering is one of the most popular fuzzy methods. FCM allows one point of data to belong to two or more clusters with different membership degrees \cite{27_arora1998approximation}. FCM was developed by Dunn in \cite{4_zhang2010neutrosophic} and improved by Bezdek \cite{5_bezdek1981objective}. FCM has four major problems: 1. It just attempts to minimize intra-cluster variance as well, but does not consider the inter-cluster variance, like k-means algorithm. 2. The result of clustering strongly depends on initializing. 3. It sensitives to noise and the membership of noise points can be high. 4. It is also sensitive to the type of distance metric and cannot distinguish between equally highly likely and equally highly unlikely data points \cite{8_menard2000fuzzy,9_yang2008alpha,10_yu2004analysis}. To solve the last problem, Gustafson and Kessel considered the Mahalanobis distance to show the shape effect on distance metric \cite{11_gustafson1979fuzzy}. In \cite{12_krishnapuram1993possibilistic}, a new method based on possibility named as possibilistic c-means (PCM) was proposed. However, it is sensitive to cluster center initialization, and needs tuning additional parameters, and may lead to generate coincident clusters. To overcome PCM problems, Pal et al. combined PCM and FCM where both the relative and absolute resemblances are considered for cluster centers \cite{13_pal1997mixed}. In \cite{14_roubens1978pattern}, fuzzy non-metric model (FNM) was proposed as a clustering approach. Richards et al. presented a variation of fuzzy and hard clustering named as relational fuzzy c-means (RFCM) \cite{15_hathaway1989relational}. In \cite{16_sen1998clustering}, Dave combined FNM and RFCM which was robust against noise and outliers. More recent researches for fuzzy c-means are evidential c-means (ECM) \cite{17_masson2008ecm} and relational evidential c-means (RECM) \cite{18_masson2009recm}. Recently, many clustering methods have been developed based on different theories \cite{Clust1li2018clustering,Clust2cui2018subspace,Clust3tong2018efficient,Clust4hammou2018convexity,Clust5saxena2017review}.

\par Neutrosophy theory was firstly proposed by Smarandache in 1995 \cite{26_smarandache1995neutrosophic}. It is a branch of philosophy, and studies the origin, nature and scope of neutralities, as well as their interactions with different ideational spectra \cite{1_smarandache2003unifying}. This theory was applied for image processing first by Guo et. al \cite{2-29_guo2009new} and then it has been successfully used for other image processing domains including image segmentation \cite{3_zhang2010neutrosophic,2-29_guo2009new,30_sengur2011color,31_heshmati2016scheme,32_guo2017efficient,salafian2018automatic}, image thresholding \cite{33_guo2014novel}, image edge detection \cite{34_guo2014novel}, image retrieval \cite{rashno2017refined,46_rashno2017content}, retinal image analysis \cite{35_rashno2017fully,36_rashno2017fully,37_rashno2017automated,38_parhi2017automated,kohler2017correlation,39_guo2017novel,40_guo2017retinal}, liver image analysis\cite{41_siri2017combined,42_sirinovel}, breast ultrasound image analysis\cite{43_lotfollahi2017segmentation}, data classification\cite{44_akbulut2017ns} and uncertainty handling\cite{45_dhar2017accurate}.  Recently, NS has been adapted for our problem of interest, data and image clustering, as neutrosophic c-means (NCM) \cite{19_guo2015ncm} and kernel neutrosophic c-means (KNCM) \cite{20_akbulut2017kncm}.

{\color{black}
\par The main motivation of this work is to handle boundary and outlier points by proposing indeterminacy set(I) in NS domain followed by presenting this set for a new clustering cost function. As it will be discussed in section 4 and 5, challenges of the previous methods are addressed  by encoding all constraints for handling boundary and outlier points in the cost function. }In this paper, we introduce a new method based on neutrosophic set (NS) theory for data clustering and image segmentation. It calculates the indeterminacy for each data point in NS domain followed by a new cost function based on data indeterminacy.  In the proposed cost function two conditions for data point \textit{i} are considered to have the highest membership degree to the main cluster k: (a) point \textit{i} should have the minimum distance from the cluster center k rather than other clusters, (b) point \textit{i} should have a small indeterminacy. Similarly, there are also two conditions for point \textit{i} to have the highest membership degree to noisy cluster: (a) having the maximum sum distance from all main clusters and (b) having a big indeterminacy. In the third step, the proposed cost function is minimized by gradient descend methods. Data points are clustered based on their membership degrees. Outlier points are assigned to noise cluster; and boundary points are assigned to main clusters with almost same membership degrees. The proposed cost function is minimized and assigns membership degrees to main and noise clusters. Here, membership sets T and F in NS domain are considered as the main and noisy clusters, respectively. The rest of the paper is organized as follows. FCM algorithm and NS set are reviewed in section 2. In section 3, the proposed method is presented. Experimental results of the proposed method in scatter and image datasets are illustrated in Section 4. Section 5 discusses advantages and disadvantages of the proposed method in comparison with other methods.  Finally, section 6 concludes the paper.

%

\section{Review on neutrosophic set and fuzzy clustering}

The proposed methods in this paper are based on NS and fuzzy clustering. In this section, these concepts are introduced as follows. 
\subsection{Neutrosophic set}
NS is a powerful framework of neutrosophy in which neutrosophic operations are defined from a technical point of view. In fact, for each application, neutrosophic sets are defined as well as neutrosophic operations corresponding to that application. Generally, in neutrosophic set $A$, each member $x$ in $A$ is denoted by three real subsets true, false and indeterminacy in interval $[0, 1]$ referred as $T$, $F$ and $I$, respectively. Each element is expressed as $x(t, i, f)$ which means that it is $t$\% true, $i$\% indeterminacy, and $f$\% false. In each application, domain experts propose the concepts behind true, false and indeterminacy\cite{2-29_guo2009new}.

\subsection{Fuzzy clustering}
 Clustering methods can classify similar samples into the same group. Consider $X$ be a data set, and xi be a sample. The purpose of clustering is to find partitions $C = \{C_1,C_2,...,C_m\},$ which satisfies \eqref{ClustCond}: 
\begin{equation}
X=\sum_{i=1}^{m}C_i\     
,C_i\neq\emptyset\ for   ( i=1,2,…,m)
,C_i\cap C_j = \emptyset\ for (i,j=1,2,…,m) ;i\neq j 
\label{ClustCond}  
\end{equation}

FCM attempts to cluster a limited number of elements $X = \{X_1,X_2,...,X_n\},$ into a collection of \textit{c} fuzzy clusters based on the similar features. Given a finite set of data, the FCM returns a list of \textit{c} cluster centers $C = \{C_1,C_2,...,C_c\},$ and a partition matrix $W=\{w_{i,j}|w_{i,j}\in[0,1],i=1,2,...,n, j=1,2,...,c\}$ where $w_{i,j}$ represents the membership degree of data point $x_i$ to cluster $c_j$.

The FCM aims to minimize objective function in \eqref{FCMObjFunc}:
\begin{equation}
	J= arg\space min \sum_{i=1}^{n}\sum_{j=1}^{c}||X_i-C_j||^2
	\label{FCMObjFunc}
\end{equation}
where membership degrees $w_{i,j}$ and  cluster centers $c_j$ are updated in each iteration by \eqref{MemDegUpd}-\eqref{ClusCentUpd}:
\begin{equation}
	w_{i,j}=\dfrac{1}{\sum_{j=1}^{c}(\dfrac{||X_i-C_j||}{||X_i-C_j||})^{2/(m-1)}}
	\label{MemDegUpd}
\end{equation}
\begin{equation}
	c_k=\dfrac{\sum_{x}w_k (x^m) x}{\sum_{x}w_k (x^m)}
	\label{ClusCentUpd}
\end{equation}
The iteration will not stop until max $\{|w_{i,j}^{(k+1)}-w_{i,j}^{(k)}|\}\leqslant\varepsilon$ where $\varepsilon$ is a small quantity and $k$ is the iteration step. This procedure attempts to achieve a minimum or a saddle point of $j$. Each data is assigned into all classes with different membership degrees\cite{FCM_bezdek1984fcm}.

\section{Proposed method}
In this paper, a new clustering approach is proposed to cluster data including outlier and boundary data points. The proposed method is derived from FCM and NS concepts. Here, uncertainty is defined for each data point and described using the indeterminacy set in neutrosophic domain.  Indeterminacy value for each data point \textit{i} is defined by considering  the Euclidean distance of \textit{i} from its neighbors by \eqref{IndDef}-\eqref{IndDef2}. 
\begin{eqnarray}
I(i)= 
\begin{cases}
1-\dfrac{NP(i)}{N/NC},&     \text{if } NP(i)< NP_{th}\\
\alpha,              &      \text{otherwise}
\end{cases}
\label{IndDef}
\end{eqnarray}

\begin{eqnarray}
temp[j]= 
\begin{cases}
1,&     \text{if } dist[i,j]<Eps, j=1,2,...,N\\
0,              &      \text{otherwise}
\end{cases}
\label{IndDef1}
\end{eqnarray}
\begin{equation}
	NP(i)=\sum_{j=1}^{N}temp[j]
	\label{IndDef2}
\end{equation}
where $I$ is an indeterminacy value, $N$ is the size of dataset, $NC$ is the number of clusters and $NP_{th}$ is a constant number as a threshold value. For indeterminacy assessment, the value of $NP_i$ is compared with $NP_{th}$. If $NP$ is smaller than $NP_{th}$, it means that this point is a noisy point and should have a bigger indeterminacy.  Otherwise, small quantity $Eps$ is considered for indeterminacy. $dist[i,j]$ is the Euclidean distance between point $i$ and $j$. It is clear that this idea assigns indeterminacy near to $1$ for noisy pixels and near to $0$ for points inside the main clusters.
\par In the proposed clustering algorithm, we consider both determinate and indeterminate membership degrees for main clusters and noisy cluster, respectively. A unique set $A$ has been considered as a union of determinate and indeterminate clusters. Let $A=C_i \cup R; i=1, 2,..., k$; where $C_i$ and  $R$ represents determinate cluster $i$ and indeterminate cluster, respectuvely.  $\cup$ is the union operation. Considering indeterminacy in clustering, the proposed objective function is defined in \eqref{ProCostFunc}: 
\begin{equation}
	L(T,F,C)=\sum_{i=1}^{n}\sum_{j=1}^{k}(w_1 I_i T_{i,j})^m ||X_i-C_j||^2 + \sum_{j=1}^{k}(w_2 (1-I_i) F_i)^m (k-||X_i-C_j||^2)
	\label{ProCostFunc}
\end{equation}
where $k$ is the number of clusters. $T_{i,j}$ and $F_i$ are the membership degrees of data point $i$ to main cluster $j$ and noisy cluster, respectively. To consider constraints in NS theory, membership degrees are enforced to be in interval  $ 0 < T_{i,j} , F_i<1$. For each data point, the sum of $T_{i,j}$ and $F_i$ should be equal to $1$ which is satisfied as following: 
\begin{equation}
	\sum_{j=1}^{k}T_{i,j} + F_i=1
	\label{MemCons}
\end{equation}
We consider two conditions for data point $i$ to have the highest membership degree to cluster $k$: $(a)$ point $i$ should have the minimum distance from the cluster center $k$ rather than other clusters, $(b)$ point $i$ should have a small indeterminacy. Similarly, there are also two conditions for point $i$ to have the highest membership degree to noisy cluster: $(a)$ having the maximum sum distance from all main clusters $\sum_{j=1}^{k}||X_i-C_j||^2$ and $(b)$ having a big indeterminacy. The maximum difference between any two pixels is $1$ since all sets have been normalized to the interval $[0, 1]$. Therefore, the maximum quantity for $\sum_{j=1}^{k}||X_i-C_j||^2$ is $k$.
\par For considering this constraint, the Lagrange function is constructed by \eqref{LagCons}:
\begin{equation}
\begin{split}
L(T,F,C)=\sum_{i=1}^{n}\sum_{j=1}^{k}(w_1 I_i T_{i,j})^m ||X_i-C_j||^2 + \sum_{j=1}^{k}(w_2 (1-I_i) F_i)^m (k-||X_i-C_j||^2)\\
-\sum_{i=1}^{n}\lambda_i (\sum_{j=1}^{k}T_{i,j} + F_i-1)
\end{split}
\label{LagCons}
\end{equation}
For cost function minimization, gradient descent approach is used. Therefore:
\begin{equation}
	\dfrac{\partial L}{\partial T_{ij}} = m(w_1 I_i T_{ij})^{(m-1)} ||X_i-C_j||^2 - \lambda_i
\end{equation}

\begin{equation}
\dfrac{\partial L}{\partial F_i} = m(w_2 (1-I_i) F_i)^{(m-1)} (K - ||X_i-C_j||^2 )- \lambda_i
\end{equation}
\begin{equation}
\dfrac{\partial L}{\partial C_j} = -(w_1 I_i T_{ij})^{m} (X_i - C_j) + (w_2 (1-I_i)F_i)^m (X_i - C_j)
\end{equation}
By considering $\dfrac{\partial L}{\partial F_i} =0$,  $\dfrac{\partial L}{\partial F_i} =0$ and $\dfrac{\partial L}{\partial C_j} =0$ we can obtain the follows:
\begin{equation}
	T_{ij}=\dfrac{1}{w_1 I_i}(\dfrac{\lambda_i}{m})^{\dfrac{1}{(m-1)}} ||X_{ij}-C_j||^{\dfrac{-2}{(m-1)}}
\end{equation}
\begin{equation}
F_i=\dfrac{1}{w_2 (1-I_i)}(\dfrac{\lambda_i}{m})^{\dfrac{1}{(m-1)}} (K-(||X_{ij}-C_j||)^2)^{\dfrac{-1}{(m-1)}}
\end{equation}
\begin{equation}
	C_j= \dfrac{[\sum_{i=1}^{n}(w_1 I_i T_{ij})^m - \sum_{i=1}^{n}(w_2 (1-I_i)F_i)]X_i}{\sum_{i=1}^{n}(w_1 I_i T_{ij})^m - \sum_{i=1}^{n}(w_2 (1-I_i)F_i)}
	\label{CompC}
\end{equation}

For efficient computation, term $(\dfrac{\lambda_i}{m})^{\dfrac{1}{m-1}}$ can be assumed as $K_{temp}$ and computed by replacing $T_{ij}$ and $F_i$ in \eqref{MemCons}:  

\begin{equation}
	(\dfrac{\lambda_i}{m})^{\dfrac{1}{m-1}} = K_{temp}
\end{equation}

\begin{equation}
\sum_{j=1}^{k}\dfrac{K_{temp}}{w_1 I_i}||X_{ij}-C_j||^{\dfrac{-2}{(m-1)}} + \dfrac{K_temp}{w_2 (1-I_i)} (K-\sum_{j=1}^{k}||X_{ij}-C_j||)^{\dfrac{-1}{(m-1)}} = 1
\end{equation}
\begin{equation}
K_{temp} = [\dfrac{1}{w_1 I_i}\sum_{j=1}^{k} ||X_{ij}-C_j||^{\dfrac{-2}{(m-1)}} + \dfrac{1}{w_2 (1-I_i)} (K-(\sum_{j=1}^{k}||X_{ij}-C_j||^2))^{\dfrac{-1}{(m-1)}}]^{-1}
\end{equation}
\begin{equation}
T_{ij}=\dfrac{K_{temp}}{w_1 I_i} ||X_{ij}-C_j||^{\dfrac{-2}{(m-1)}}
\label{CompT}
\end{equation}
\begin{equation}
F_i=\dfrac{K_{temp}}{w_2 (1-I_i)} (K-(\sum_{j=1}^{k}||X_{ij}-C_j||^2))^{\dfrac{-1}{(m-1)}}
\label{CompF}
\end{equation}
The proposed clustering algorithm is summarized as follows:
\begin{algorithm}
	\caption{}\label{euclid}
	\begin{algorithmic}[1]
		\State Initialize $T$ and $F$.
		\State Initialize the $c$, $w1$, $w2$, $Eps$, $K$ and $m$.
		\State Compute $I$ for each data point.
		\State Update $T_{ij}$, $F_{i}$ and $C_i$ by Eqs \eqref{CompT}, \eqref{CompF} and \eqref{CompC}, respectively.
		\State Check the stop condition, if $|T^{(k)} - T^{(k-1)}| < \varepsilon$ then stop, otherwise go to $step$ $4$.
		\State Assign each data point into boundary cluster if the first two membership degrees $T_{ij}$ and $T_{ik}$ are between $t$ and $(1-t)$, otherwise assign it to a cluster  which data point $i$ has the maximum membership degree to it.
		\State end.
	\end{algorithmic}
\end{algorithm}
\section{Experimental Results }
Performance of the proposed clustering method is evaluated in three types of datasets including diamond datasets, natural and artificial images datasets and medical image dataset in sections 4.1, 4.2 and 4.3, respectively. The proposed method is applied on these datasets and then compared with ASIC \cite{48_pal1991fuzzy}, FCM-AWA \cite{49_kang2009novel}, NCM \cite{19_guo2015ncm} and methods in \cite{35_rashno2017fully,36_rashno2017fully,51_boykov2006graph,52_salah2011multiregion,53_esmaeili2016three,54_de2015machine,55_venhuizena2015vendor}. 
\subsection{Parameters Tuning}
All parameters in indeterminacy computation section are set to following quantities based on experiments. Parameter $Eps$ and $NP_{th}$ are considered with quantity $4$, means that neighbors in the maximum distance of $4$ and the maximum of $4$ neighbors are considered for indeterminacy computation. Parameter $\alpha$ is considered with quantity $0.05$. In the proposed cost function, parameter are configured as $\varepsilon=10 ^{-6}$, $m=2$, $t=0.4$, $w_1=1$ and $w_2=2$.

{\color{black}
\subsection{Datasets}
In this research, three type of datasets are used to evaluate the performance of the proposed method. The first type is diamond datasets including a collections of scatter data including X12, X19 and X24 which are proposed in \cite{19_guo2015ncm}. We also proposed X35 which is an extension of datasets in \cite{19_guo2015ncm}. In these datasets, boundary points are considered between main clusters as well as outlier points far from main clusters. It can be visually seen that how clustering methods are affected by these points among main points in each dataset. The second dataset type is UCI which include datasets with higher dimension and larger scale. Here, "Iris", "Wine", "Glass", "Seeds" and "Breast-w" are used. Finally, the proposed method is applied on image data as third dataset type. In this experiment, these dataset natural, artificial and medical images are used to evaluate the effectiveness of the proposed method.   
}
\subsection{Diamond datasets}
Diamond datasets in NCM including: $a)$ $X12: 12$ scatter points in $2$ clusters with $1$ boundary point and $1$ outlier, $b)$ $X19: 19$ scatter points in $3$ clusters with $2$ boundary points and $2$ outliers and $c)$ $X24: 24$ scatter points in $4$ clusters with $3$ boundary points and $1$ outlier were used in this research. We also designed a further scatter dataset referred as $X35$.
\subsubsection{X12 dataset}
\begin{figure}[!hp]
	\centering
	\includegraphics[width= 0.7 \textwidth ,height=7cm]{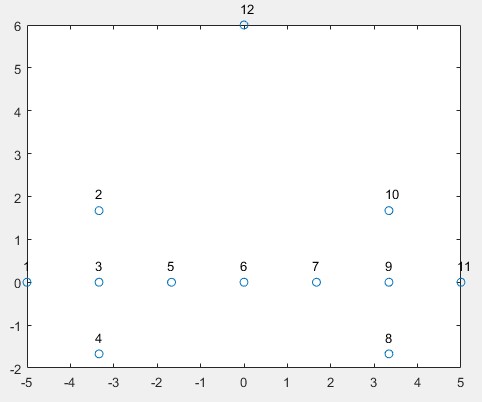}
	\caption{X12 dataset.}
	\label{X12}
\end{figure}

\begin{figure}[!hp]
	\centering
	\includegraphics[width= 0.7 \textwidth ,height=7cm]{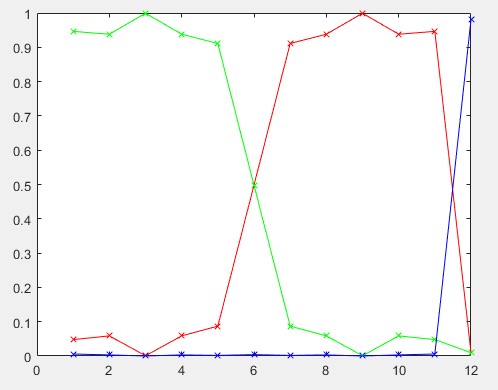}
	\caption{Membership degrees for X12 dataset computed by the proposed method.}
	\label{X12MD}
\end{figure}

\begin{figure}[!t]
	\centering
	\includegraphics[width= 0.7 \textwidth ,height=7cm]{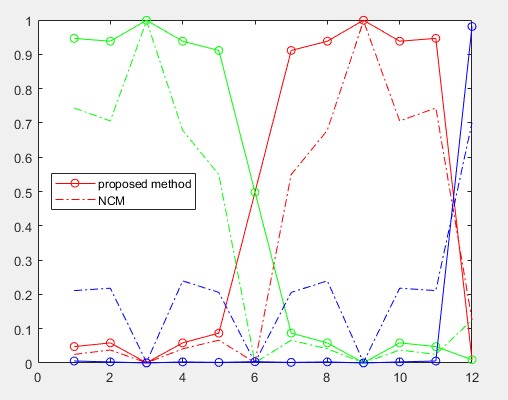}
	\caption{Membership degrees computed by NCM and the proposed method in X12.}
	\vspace{-1.5em}
	\label{X12NCMOur}
\end{figure}

$X12$ is shown in Fig. \ref{X12} in which points $1-5$ and $7-11$ belong to main clusters, points $6$ and $12$ are boundary and outlier, respectively. In all experiments, $T_{ci}$ is the membership degree to $i_{th}$ cluster and $F$ represents the membership degree to noise cluster. Each data point is assigned to a cluster with the maximum membership degree. Fig. \ref{X12MD} illustrates assigned membership degrees of each point to two main clusters and noisy cluster by the proposed method with $red$, $green$ and $blue$ colors, respectively.  Table \ref{X12Table} reports the membership degrees assigned by the proposed clustering method and NCM. Although, both the proposed method and NCM assign correct cluster labels to all points, the proposed method determines the cluster labels more confidently. For example, for point $7$ which is a point in cluster $2$, the proposed method assigns $0.91$ membership degree to this cluster while NCM assigns $0.69$. All membership degreess are also visually depicted in Fig. \ref{X12NCMOur} . Dash $("-.-.-")$ and circle $("o-o-o")$ represent membership degrees computed by NCM and the proposed method, respectively.

\begin{table}[]
	\centering
	\caption{Clustering result for X12 dataset.}
	\label{X12Table}
	\resizebox{\linewidth}{4cm}{
	\begin{tabular}{|l|l|l|l|l|l|l|l|l|}
			\hline
		& \multicolumn{4}{|c|}{NCM}           &         \multicolumn{4}{c|}{Proposed method} \\ \hline
		& Tc1    & Tc2    & I      & F      & Tc1    & Tc2    & F      &     \\ \hline
		1 & 0.8262 & 0.0294 & 0.0104 & 0.1339 & 0.9515 & 0.0479 & 0.0006 &          \\ \hline 
		2 & 0.7952 & 0.0451 & 0.0196 & 0.1401 & 0.9409 & 0.0588 & 0.0003 &          \\ \hline
		3 & 0.9996 & 0.0001 & 0.0000 & 0.0003 & 0.9993 & 0.0007 & 0.0000 &          \\ \hline
		4 & 0.792  & 0.0456 & 0.0197 & 0.1426 & 0.9409 & 0.0588 & 0.0003 &          \\ \hline
		5 & 0.695  & 0.0799 & 0.0915 & 0.1336 & 0.9124 & 0.0874 & 0.0002 &          \\ \hline
		6 & 0.0007 & 0.0007 & 0.9982 & 0.0005 & 0.4998 & 0.4998 & 0.0004 & \textbf{boundary} \\ \hline
		7 & 0.0835 & 0.6802 & 0.0990 & 0.1373 & 0.0874 & 0.9124 & 0.0002 &          \\ \hline
		8 & 0.0475 & 0.7854 & 0.0207 & 0.1464 & 0.0588 & 0.9409 & 0.0003 &          \\ \hline
		9 & 0.0003 & 0.9987 & 0.0001 & 0.0008 & 0.0007 & 0.9993 & 0.0000 &          \\ \hline
		10 & 0.0444 & 0.7994 & 0.0195 & 0.1367 & 0.0588 & 0.9409 & 0.0003 &          \\ \hline
		11 & 0.0284 & 0.8334 & 0.0101 & 0.128  & 0.0479 & 0.9515 & 0.0006 &          \\ \hline
		12 & 0.0477 & 0.0938 & 0.0084 & 0.8502 & 0.0765 & 0.0765 & 0.847  &       \\ \hline
	\end{tabular}
}
\end{table}
\subsubsection{X19 datasets}
\begin{figure}[!t]
	\centering
	\includegraphics[width= 0.7 \textwidth ,height=7cm]{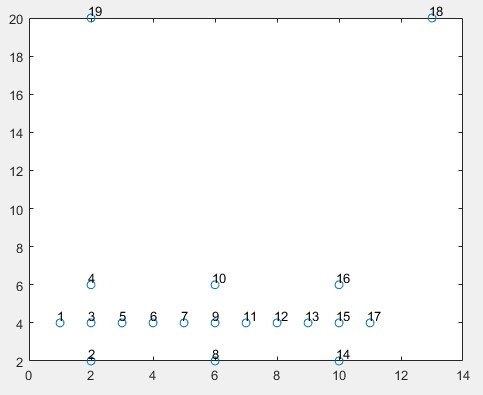}
	\caption{X19 dataset.}
	\label{X19}
\end{figure}

\begin{figure}[!t]
	\centering
	\includegraphics[width= 0.7 \textwidth ,height=7cm]{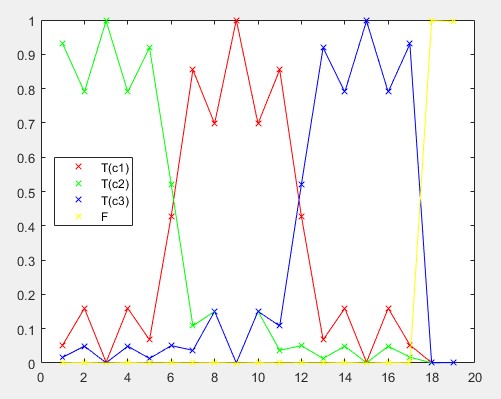}
	\caption{Membership degrees computed by the proposed method in X19.}
	\vspace{-0.05em}
	\label{X19MD}
\end{figure}

$X19$ dataset with three clusters is shown in Fig. \ref{X19} . In this dataset, points $1-5$, $7-11$ and $13-17$ represent main clusters, points $6$ and $12$ are boundary and points $18$ and $19$ are noisy points. Membership degrees computed by the proposed method and NCM are reported in Table \ref{X19Table} . Similar to section $4.2.1$, although the proposed method and NCM assign same cluster labels for all points, the proposed method assigns membership degrees for points such as $5$, $7$, $11$, $12$ with  higher certainties into their corresponding clusters. Point $6$ belongs to boundary cluster and point $3$ is a main cluster center. Although, point $5$ has a same distance between main and boundary clusters, it belongs to main cluster. NCM cannot distinguish boundary and main clusters for point $5$. The proposed method addresses this issue and assigns $0.91$ of membership degree to main cluster while NCM assigns $0.58$. Fig. \ref{X19MD} depicts membership degrees visually.
\begin{table}[]
	\centering
	\caption{Clustering result for X19 dataset.}
	\label{X19Table}
	\resizebox{\linewidth}{5cm}{
	\begin{tabular}{|l|l|l|l|l|l|l|l|l|l|l|}
		\hline
		& \multicolumn{5}{c|}{NCM}                    & \multicolumn{5}{c|}{Proposed method}          \\ \hline
		& Tc1    & Tc2    & Tc3    & I      & F      & T1     & T2     & T3     & F      &     \\ \hline
		1  & 0.89   & 0.0236 & 0.007  & 0.015  & 0.0645 & 0.0169 & 0.9307 & 0.0525 & 0      &          \\	\hline
		2  & 0.7759 & 0.0578 & 0.0144 & 0.0444 & 0.1076 & 0.0491 & 0.7913 & 0.1596 & 0      &          \\	\hline
		3  & 0.988  & 0.003  & 0.0007 & 0.0029 & 0.0053 & 0.0006 & 0.9972 & 0.0022 & 0      &          \\	\hline
		4  & 0.8393 & 0.0411 & 0.0103 & 0.0332 & 0.0762 & 0.0491 & 0.7913 & 0.1596 & 0      &          \\	\hline
		5  & 0.5816 & 0.0928 & 0.0161 & 0.2182 & 0.0913 & 0.0131 & 0.9194 & 0.0675 & 0      &          \\	\hline
		6  & 0.0124 & 0.0149 & 0.0016 & 0.9646 & 0.0065 & 0.0507 & 0.5211 & 0.4282 & 0      & \textbf{boundary} \\	\hline
		7  & 0.0689 & 0.7032 & 0.0261 & 0.1249 & 0.0769 & 0.0369 & 0.1081 & 0.855  & 0      &          \\	\hline
		8  & 0.0434 & 0.792  & 0.0434 & 0.0317 & 0.0894 & 0.1507 & 0.1507 & 0.6985 & 0      &          \\	\hline
		9  & 0      & 0.9999 & 0      & 0      & 0      & 0      & 0      & 1      & 0      &          \\	\hline
		10 & 0.0421 & 0.7996 & 0.0421 & 0.0315 & 0.0847 & 0.1507 & 0.1507 & 0.6985 & 0      &          \\	\hline
		11 & 0.0261 & 0.7032 & 0.0689 & 0.1249 & 0.0769 & 0.1081 & 0.0369 & 0.855  & 0      &          \\	\hline
		12 & 0.0016 & 0.0149 & 0.0124 & 0.9646 & 0.0065 & 0.5211 & 0.0507 & 0.4282 & 0      & \textbf{boundary} \\	\hline
		13 & 0.0161 & 0.0928 & 0.5816 & 0.2182 & 0.0913 & 0.9194 & 0.0131 & 0.0675 & 0      &          \\	\hline
		14 & 0.0144 & 0.0578 & 0.7759 & 0.0444 & 0.1076 & 0.7913 & 0.0491 & 0.1596 & 0      &          \\	\hline
		15 & 0.0007 & 0.003  & 0.988  & 0.0029 & 0.0053 & 0.9972 & 0.0006 & 0.0022 & 0      &          \\	\hline
		16 & 0.0103 & 0.0411 & 0.8393 & 0.0332 & 0.0762 & 0.7913 & 0.0491 & 0.1596 & 0      &          \\	\hline
		17 & 0.007  & 0.0236 & 0.89   & 0.015  & 0.0645 & 0.9307 & 0.0169 & 0.0525 & 0      &          \\	\hline
		18 & 0.037  & 0.0854 & 0.3046 & 0.0324 & 0.5406 & 0.0458 & 0.0327 & 0.0399 & 0.8817 &          \\	\hline
		19 & 0.3046 & 0.0854 & 0.037  & 0.0324 & 0.5406 & 0.0616 & 0.0763 & 0.0718 & 0.7904 &          \\ \hline
	\end{tabular}
}
\end{table}
\subsubsection{X24 dataset}
\begin{figure}[!hp]
	\centering
	\includegraphics[width= 0.7 \textwidth ,height=7cm]{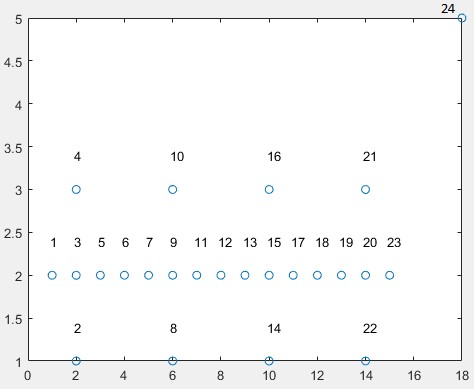}
	\caption{X24 dataset.}
	\vspace{2em}
	\label{X24}
\end{figure}

\begin{figure}[!hp]
	\centering
	\includegraphics[width= 0.7 \textwidth ,height=7cm]{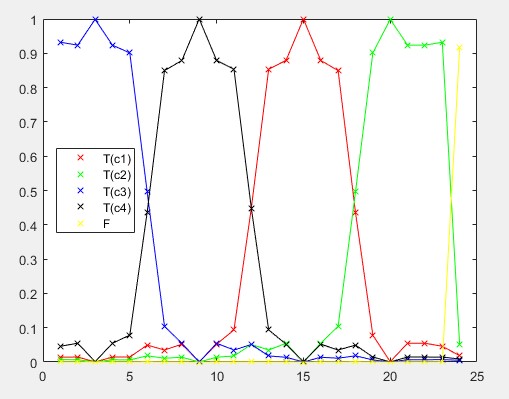}
	\caption{Membership degrees computed by the proposed method in X24.}
	\vspace{-1.5em}
	\label{X24MD}
\end{figure}

We also conducted more experiment to compare the proposed method and NCM using a four class dataset $X24$ which is represented in Fig. \ref{X24} . Data points $6$, $12$ and $18$ are boundary and $24$ is an outlier. The results of the proposed method and NCM are tabulated in Table \ref{X24Table} . The first five data points belong to the first main cluster because of their higher $T_{c1}$ values. Similar observation can be inferred for the other clusters ($C2$ and $C3$ and $C4$). Data points $6$, $12$ and $18$ are ambiguous because of the two highest $T$ values. Finally, last data point $(24)$ is deduced as outlier.  Fig. \ref{X24MD} depicts membership degrees visually.
\begin{table}[!hp]
	\centering
	\caption{Clustering result for X24 dataset.} 
	\label{X24Table}
	\resizebox{\linewidth}{5cm}{
	\begin{tabular}{|l|l|l|l|l|l|l|l|l|l|l|l|l|}
		\hline
		&  \multicolumn{6}{c|}{NCM}                             & \multicolumn{6}{c|}{Proposed method}                   \\ \hline
		& Tc1    & Tc2    & Tc3    & Tc4    & I      & F      & Tc1    & Tc2    & Tc3    & Tc4    & F      &     \\ \hline
		1  & 0.0248 & 0.0075 & 0.8631 & 0.0035 & 0.0363 & 0.0648 & 0.0459 & 0.0069 & 0.9322 & 0.0143 & 0.0006 &          \\ \hline
		2  & 0.0464 & 0.0119 & 0.77   & 0.0052 & 0.0839 & 0.0825 & 0.0547 & 0.0066 & 0.9238 & 0.0144 & 0.0004 &          \\ \hline
		3  & 0.0015 & 0.0004 & 0.9925 & 0.0002 & 0.0031 & 0.0024 & 0.0008 & 0.0001 & 0.9989 & 0.0002 & 0      &          \\ \hline
		4  & 0.0464 & 0.0119 & 0.7703 & 0.0052 & 0.0839 & 0.0824 & 0.0547 & 0.0066 & 0.9238 & 0.0144 & 0.0004 &          \\ \hline
		5  & 0.0703 & 0.0126 & 0.4752 & 0.005  & 0.3709 & 0.0661 & 0.0782 & 0.006  & 0.901  & 0.0145 & 0.0003 &          \\ \hline
		6  & 0.003  & 0.0003 & 0.0026 & 0.0001 & 0.9928 & 0.0012 & 0.4364 & 0.0181 & 0.4957 & 0.0492 & 0.0007 & \textbf{boundary} \\ \hline
		7  & 0.5937 & 0.023  & 0.0595 & 0.0069 & 0.255  & 0.0619 & 0.8495 & 0.011  & 0.1043 & 0.0349 & 0.0003 &          \\ \hline
		8  & 0.7527 & 0.0428 & 0.0414 & 0.0108 & 0.0732 & 0.0791 & 0.8792 & 0.0139 & 0.0546 & 0.052  & 0.0003 &          \\ \hline
		9  & 1      & 0      & 0      & 0      & 0      & 0      & 1      & 0      & 0      & 0      & 0      &          \\ \hline
		10 & 0.7537 & 0.0427 & 0.0412 & 0.0108 & 0.073  & 0.0786 & 0.8792 & 0.0139 & 0.0546 & 0.052  & 0.0003 &          \\ \hline
		11 & 0.5769 & 0.0611 & 0.0218 & 0.0109 & 0.2687 & 0.0606 & 0.8538 & 0.0176 & 0.0349 & 0.0934 & 0.0003 &          \\ \hline
		12 & 0.0007 & 0.0007 & 0.0001 & 0.0001 & 0.9981 & 0.0003 & 0.4485 & 0.0512 & 0.0512 & 0.4485 & 0.0006 & \textbf{boundary} \\ \hline
		13 & 0.0663 & 0.5152 & 0.0117 & 0.0215 & 0.3228 & 0.0626 & 0.0934 & 0.0349 & 0.0176 & 0.8538 & 0.0003 &          \\ \hline
		14 & 0.0449 & 0.7462 & 0.0113 & 0.0392 & 0.0785 & 0.0799 & 0.052  & 0.0546 & 0.0139 & 0.8792 & 0.0003 &          \\ \hline
		15 & 0.0004 & 0.9978 & 0.0001 & 0.0003 & 0.0008 & 0.0006 & 0      & 0      & 0      & 1      & 0      &          \\ \hline
		16 & 0.0437 & 0.7518 & 0.011  & 0.0389 & 0.0768 & 0.0778 & 0.052  & 0.0546 & 0.0139 & 0.8792 & 0.0003 &          \\ \hline
		17 & 0.0224 & 0.6581 & 0.0067 & 0.0522 & 0.202  & 0.0587 & 0.0349 & 0.1043 & 0.011  & 0.8495 & 0.0003 &          \\ \hline
		18 & 0.0018 & 0.0174 & 0.0006 & 0.0124 & 0.9611 & 0.0067 & 0.0492 & 0.4957 & 0.0181 & 0.4364 & 0.0007 & \textbf{boundary} \\ \hline
		19 & 0.0128 & 0.0733 & 0.005  & 0.3824 & 0.461  & 0.0655 & 0.0145 & 0.901  & 0.006  & 0.0782 & 0.0003 &          \\ \hline
		20 & 0.0013 & 0.0053 & 0.0006 & 0.9722 & 0.0122 & 0.0085 & 0.0002 & 0.9989 & 0.0001 & 0.0008 & 0      &          \\ \hline
		21 & 0.011  & 0.0438 & 0.0048 & 0.7807 & 0.0846 & 0.075  & 0.0144 & 0.9238 & 0.0066 & 0.0547 & 0.0004 &          \\ \hline
		22 & 0.014  & 0.0553 & 0.0061 & 0.7268 & 0.1028 & 0.095  & 0.0144 & 0.9238 & 0.0066 & 0.0547 & 0.0004 &          \\ \hline
		23 & 0.0058 & 0.0195 & 0.0027 & 0.8928 & 0.0297 & 0.0495 & 0.0143 & 0.9322 & 0.0069 & 0.0459 & 0.0006 &          \\ \hline
		24 & 0.0353 & 0.0753 & 0.02   & 0.2413 & 0.0633 & 0.5649 & 0.0087 & 0.0515 & 0.0051 & 0.0182 & 0.9165 &       \\ \hline  
	\end{tabular}
}
\end{table}
\subsubsection{X35 dataset}
\begin{figure}[H]
	\centering
	\includegraphics[width= 0.7 \textwidth ,height=7cm]{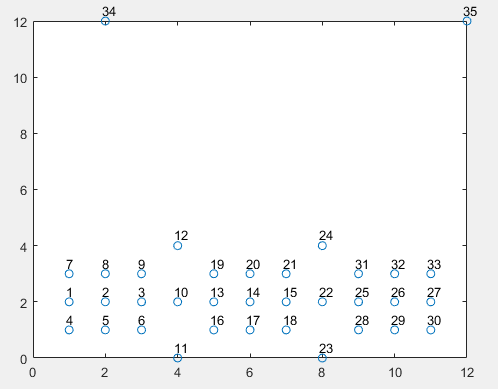}
	\caption{X35 dataset.}
	\vspace{-0.5em}
	\label{X35}
\end{figure}

\begin{figure}[H]
	\centering
	\includegraphics[width= 0.7 \textwidth ,height=7cm]{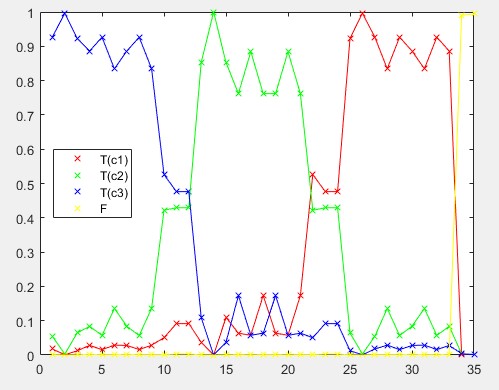}
	\caption{Membership degrees computed by the proposed method in X35.}
	\vspace{-1.5em}
	\label{X35MD}
\end{figure}

We also evaluated the proposed method for our diamond dataset with $35$ data points which is shown in Fig. \ref{X35} . Points $1-9$, $13-21$ and $25-33$ belong to main clusters, points $10-12$ and $22-24$ are ambiguous and points $34$ and $35$ are outliers. In this dataset, we have considered more ambiguous points to see the effect of boundary points in final clustering results. Table \ref{X35Table} reports membership degrees computed by the proposed method and NCM. In NCM, points $(3, 6, 9)$, $(13, 16, 19)$, $(15, 18, 21)$ and $(25, 28, 31)$ are assigned to main clusters with almost $0.50$ membership degree, while in the proposed method, these points are assigned to main clusters with membership degree between $0.80$ to $0.90$. In fact, the proposed clustering scheme can solve the clustering problems in boundary points more efficiently. The membership degrees for each data point are visually depicted in Fig. \ref{X35MD} . 
\begin{table}[H]
	\centering
	\caption{Clustering result for X35 dataset.}
	\label{X35Table}
	\resizebox{\linewidth}{9cm}{
	\begin{tabular}{|l|l|l|l|l|l|l|l|l|l|l|}
		\hline
		& \multicolumn{5}{c|}{NCM}                    & \multicolumn{5}{c|}{Proposed method}          \\ \hline
		& Tc1    & Tc2    & Tc3    & I      & F      & Tc1    & Tc2    & Tc3    & F      &     \\ \hline
		1  & 0.0069 & 0.0234 & 0.8738 & 0.0346 & 0.0612 & 0.0175 & 0.0541 & 0.9271 & 0.0013 &          \\ \hline
		2  & 0.0005 & 0.0021 & 0.9894 & 0.0046 & 0.0035 & 0.0007 & 0.0027 & 0.9965 & 0      &          \\ \hline
		3  & 0.0125 & 0.0714 & 0.46   & 0.3888 & 0.0673 & 0.0125 & 0.0643 & 0.9227 & 0.0005 &          \\ \hline
		4  & 0.0148 & 0.0485 & 0.738  & 0.0664 & 0.1323 & 0.0278 & 0.0839 & 0.8862 & 0.0021 &          \\ \hline
		5  & 0.0121 & 0.0483 & 0.7656 & 0.0878 & 0.0862 & 0.0157 & 0.0569 & 0.9266 & 0.0009 &          \\ \hline
		6  & 0.0211 & 0.1106 & 0.4555 & 0.2965 & 0.1163 & 0.0288 & 0.1357 & 0.8343 & 0.0011 &          \\ \hline
		7  & 0.0142 & 0.0469 & 0.7471 & 0.0646 & 0.1272 & 0.0278 & 0.0839 & 0.8862 & 0.0021 &          \\ \hline
		8  & 0.0114 & 0.0455 & 0.7785 & 0.0839 & 0.0807 & 0.0157 & 0.0569 & 0.9266 & 0.0009 &          \\ \hline
		9  & 0.0205 & 0.1081 & 0.4566 & 0.3023 & 0.1125 & 0.0288 & 0.1357 & 0.8343 & 0.0011 &          \\ \hline
		10 & 0.0005 & 0.0044 & 0.0037 & 0.9896 & 0.0018 & 0.0503 & 0.4223 & 0.5261 & 0.0013 & \textbf{boundary} \\ \hline
		11 & 0.0485 & 0.2536 & 0.2324 & 0.2504 & 0.2151 & 0.0915 & 0.4294 & 0.4764 & 0.0028 & \textbf{boundary} \\ \hline
		12 & 0.0476 & 0.2543 & 0.2314 & 0.2565 & 0.2101 & 0.0915 & 0.4294 & 0.4764 & 0.0028 & \textbf{boundary} \\ \hline
		13 & 0.0224 & 0.6022 & 0.0595 & 0.2528 & 0.063  & 0.0371 & 0.8528 & 0.1094 & 0.0006 &          \\ \hline
		14 & 0      & 0.9998 & 0      & 0.0001 & 0.0001 & 0      & 1      & 0      & 0      &          \\ \hline
		15 & 0.0595 & 0.6023 & 0.0225 & 0.2527 & 0.0631 & 0.1095 & 0.8527 & 0.0372 & 0.0006 &          \\ \hline
		16 & 0.0381 & 0.5179 & 0.0952 & 0.2375 & 0.1113 & 0.0636 & 0.762  & 0.1733 & 0.0012 &          \\ \hline
		17 & 0.043  & 0.7569 & 0.043  & 0.0736 & 0.0835 & 0.0575 & 0.8843 & 0.0575 & 0.0007 &          \\ \hline
		18 & 0.0951 & 0.5181 & 0.0382 & 0.2372 & 0.1114 & 0.1733 & 0.7619 & 0.0636 & 0.0012 &          \\ \hline
		19 & 0.0367 & 0.5253 & 0.0923 & 0.2386 & 0.107  & 0.0636 & 0.762  & 0.1733 & 0.0012 &          \\ \hline
		20 & 0.0398 & 0.7747 & 0.0398 & 0.0689 & 0.0768 & 0.0575 & 0.8843 & 0.0575 & 0.0007 &          \\ \hline
		21 & 0.0922 & 0.5253 & 0.0368 & 0.2386 & 0.1071 & 0.1733 & 0.7619 & 0.0636 & 0.0012 &          \\ \hline
		22 & 0.0038 & 0.0045 & 0.0005 & 0.9894 & 0.0019 & 0.5262 & 0.4221 & 0.0503 & 0.0013 & \textbf{boundary} \\ \hline
		23 & 0.232  & 0.2537 & 0.0486 & 0.2503 & 0.2153 & 0.4764 & 0.4293 & 0.0915 & 0.0028 & \textbf{boundary} \\ \hline
		24 & 0.2313 & 0.2543 & 0.0476 & 0.2567 & 0.2101 & 0.4764 & 0.4293 & 0.0915 & 0.0028 & \textbf{boundary} \\ \hline
		25 & 0.4585 & 0.0715 & 0.0125 & 0.3902 & 0.0674 & 0.9228 & 0.0642 & 0.0125 & 0.0005 &          \\ \hline
		26 & 0.9891 & 0.0021 & 0.0005 & 0.0047 & 0.0036 & 0.9965 & 0.0027 & 0.0007 & 0      &          \\ \hline
		27 & 0.8745 & 0.0233 & 0.0069 & 0.0344 & 0.0609 & 0.927  & 0.0541 & 0.0175 & 0.0013 &          \\ \hline
		28 & 0.4545 & 0.1107 & 0.0212 & 0.297  & 0.1165 & 0.8344 & 0.1357 & 0.0288 & 0.0011 &          \\ \hline
		29 & 0.7648 & 0.0485 & 0.0122 & 0.0881 & 0.0865 & 0.9266 & 0.0569 & 0.0157 & 0.0009 &          \\ \hline
		30 & 0.7381 & 0.0485 & 0.0148 & 0.0664 & 0.1322 & 0.8862 & 0.0839 & 0.0278 & 0.0021 &          \\ \hline
		31 & 0.4561 & 0.1081 & 0.0205 & 0.3029 & 0.1125 & 0.8344 & 0.1357 & 0.0288 & 0.0011 &          \\ \hline
		32 & 0.7788 & 0.0454 & 0.0113 & 0.0838 & 0.0806 & 0.9266 & 0.0569 & 0.0157 & 0.0009 &          \\ \hline
		33 & 0.748  & 0.0467 & 0.0142 & 0.0644 & 0.1268 & 0.8862 & 0.0839 & 0.0278 & 0.0021 &          \\\hline
		34 & 0.0447 & 0.0645 & 0.0748 & 0.0361 & 0.7798 & 0.0024 & 0.0033 & 0.0038 & 0.9906 &          \\ \hline
		35 & 0.0742 & 0.0563 & 0.0375 & 0.0332 & 0.7987 & 0.0023 & 0.0018 & 0.0012 & 0.9947 &  
		\\ \hline       
	\end{tabular}
}
\end{table}

{\color{black}
\subsection{UCI Datasets}

To show the performance of the proposed method on larger scale datasets, UCI datasets are used. These datasets are considered as standard datasets in machine learning community. In this research, "Iris", "Wine", "Glass", "Seeds" and "Breast-w" datasets are selected among other datasets in UCI. Table \ref{summary} summaries number of features, number of classes, samples in each cluster and number of objects in each dataset. Accuracy of the proposed method and FCM \cite{FCM_new}, PCM \cite{PCM_new}, PFCM \cite{PFCM_new} and HPFCM \cite{HPFCM_new} methods are summarized in Table \ref{acc}. The proposed method outperforms other methods in "Iris", "Wine", "Glass", "Seeds" and Breast-w datasets with the accuracy of 94.66\%, 83.14\%, 91.58\% and 91.90\% and 91.41\%, respectively. 

\begin{table}[H]
	\centering
	\caption{Summary of dataset characteristics:}
	\label{summary}
	\begin{tabular}{lllll}
		Dataset  & No. of feature & No. of classes & No. each cluster & No. object \\ \hline
		Iris     & 4              & 3              & 50,50,50         & 150        \\ \hline
		Wine     & 13             & 3              & 48,59,71         & 178        \\ \hline
		Glass    & 9              & 6              & 9,29,13,70,17,76 & 214        \\ \hline
		Seed     & 7              & 3              & 70,70,70         & 210        \\ \hline
		Breast-w & 9              & 2              & 241,458          & 699       \\ \hline
	\end{tabular}
\end{table}

\begin{table}[H]
	\centering
	\caption{ Clustering accuracy for FCM, PCM, PFCM, HPFCM and Proposed method with five datasets: Iris, Wine, Glass, seeds and Breast cancer. }
	\label{acc}
	\begin{tabular}{|l|l|l|l|l|l|}  \hline
		Data sets & FCM  & PCM  & PFCM & HPFCM & Proposed method         \\ \hline
		Iris      & 89.3 & 66.7 & 90   & 92.9  & \textbf{94.66}  \\ \hline
		Wine      & 68.5 & 41.5 & 70   & {78.9}  & \textbf{83.14}          \\ \hline
		Glass     & 72.1 & 55.4 & 82.3 & 87.6  & \textbf{91.58} \\ \hline
		Seeds     & 78.3 &69.8  &84.3  & 86.6  & \textbf{91.90}  \\ \hline
		Breast-w  & 84.3 & 61.8 & 86.3 & 89.1  & \textbf{91.41}        \\ \hline      
	\end{tabular}
\end{table}

}

\subsection{Natural and artificial images datasets }
Pixel clustering can be used for image segmentation in which each cluster is considered as a segment. Each pixel's intensity is used as a one dimensional data for clustering algorithm. This dataset includes natural and artificial images.  In this section, we have applied the proposed method to image segmentation and compared with the existing image segmentation algorithms such as NCM, ASIC \cite{48_pal1991fuzzy} and FCM–AWA \cite{49_kang2009novel}.
\par
\begin{figure}[!t]
	\centering
	\includegraphics[width= 1 \textwidth ]{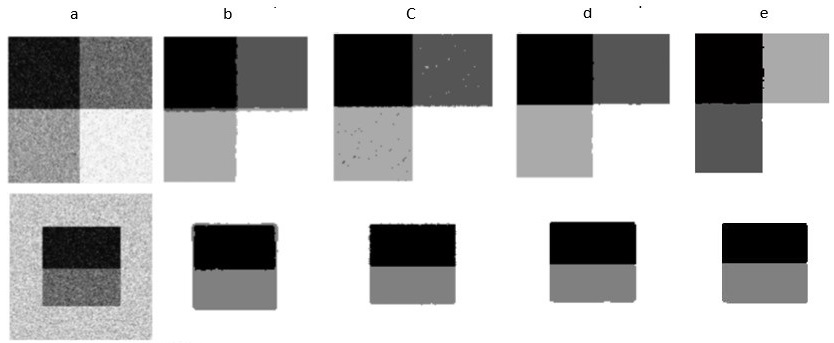}
	\caption{Segmentation results for two samples of artificial images (each sample in one row). (a) original image, semented image by (b) FCM-AWA, (c) ASIC, (d) NCM and (e) the proposed method.}
	\label{ArtImgSq}
\end{figure}

 In the proposed method, membership sets $T$ and $F$ should be post processed so they can be used for pixels clustering \cite{6-7_baraldi1999survey}. For each pixel, the average of its neighbors is calculated to descend the influence of undesired factors on the final determination of membership sets. Therefore, image clustering process is same with scatter data except final membership of each pixel is calculated by \eqref{MemImage1}-\eqref{MemImage2}:
\begin{equation}
	\overline{T}(i,j)=\dfrac{1}{z^2}\sum_{m,n\in S} T(m,n)
	\label{MemImage1}
\end{equation}
\begin{equation}
\overline{F}(i,j)=\dfrac{1}{z^2}\sum_{m,n\in S} F(m,n)
\label{MemImage2}
\end{equation}
where $z$ represents the size of $S$, which has been set to $3$ in this application. Fig. \ref{ArtImgSq} shows two artificial images in which each row contains the segmentation result of different methods  for one image. The first row in Fig. \ref{ArtImgSq} is a synthesized image with four classes and the corresponding gray values are $50$ (upper left, UL), $100$ (upper right, UR), $150$ (low left, LL) and $200$ (low right, LR), respectively. Each cluster (sub-image) contains $64 × 64$ pixels. The image is degraded by the Gaussian noise $(\mu=0$, $\sigma=25.5)$. The second row shows another synthesized image that contains three regions: two equal-sized rectangular on a uniform background and corresponding gray values $20$ (upper step), $100$ (lower step) and $255$ (background). Gaussian noise $(\mu=0$, $\sigma=25.5)$ is also added to this image.
\par It is visually clear from segmentation results that the proposed method archives good homogeneity in the segmented regions in comparison with NCM, ASIC and FCM–AWA. In the proposed method, boundaries of the homogenous regions are smooth and a few number of  pixels are misclassiﬁed. As it is reported in Table \ref{ArtImgSqTable} , the proposed method creates $(43,9)$ misclassiﬁed pixels in the segmentation of the artidicial images (Image1, Image 2) which outperforms NCM, FCM-AWA and ASIC with (70,15), (418,235) and (144,47) misclassiﬁed pixels, respectively. 
 Note that FCM-AWA has been tested by setting parameters: $m=2$, $\alpha=50$, $\varepsilon=10^{-5}$, $r=2$, $k_0=0.45$ and $k_1=0.65$. In ASIC method, the cooling factor $\alpha$ is set to $0.95$. 
\begin{table}[]
	\centering
	\caption{Number of misclassiﬁed pixels in two artificial images.}
	\label{ArtImgSqTable}
	\begin{tabular}{|l|l|l|l|l|}
		\hline
		& ASIC & FCM-AWA & NCM & Proposed Method \\
		\hline
		Image 1 & 144  & 418     & 70  & 		\textbf{43}              \\ \hline
		Image 2 & 47   & 235     & 15  & 		\textbf{9}  \\ \hline     
	\end{tabular}
\end{table}
\par
\begin{figure}[H]
	\centering
	\includegraphics[width= 1 \textwidth ]{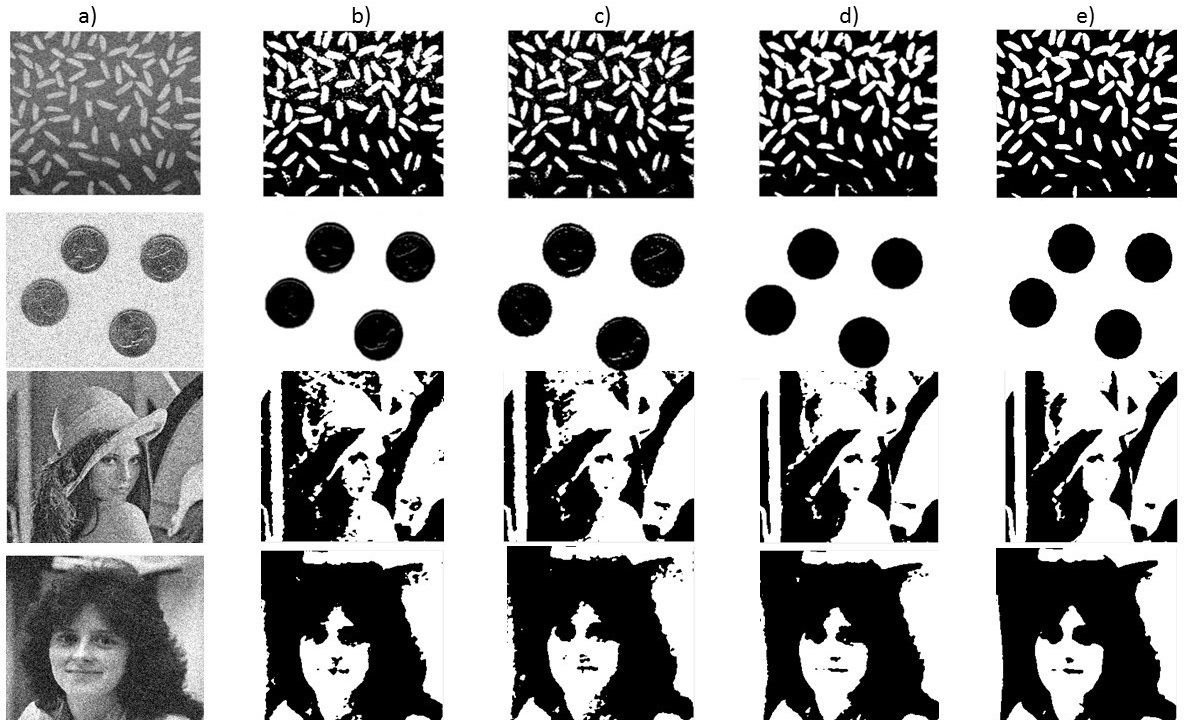}
	\caption{Segmentation results for four samples of natural images (each sample in one row). (a) original image, semented image by (b) ASCI, (c) FCM-AWA, (d) NCM and (e) the proposed method.}
	\vspace{-1.5em}
	\label{NatImg}
\end{figure}

 We also compared the proposed method with ASIC, FCM-AWA and NCM methods in four natural images: $rice$, $eight$, $Lena$ and $women$ in the first, second, third and forth rows in Fig. \ref{NatImg} , respectively. In these cases, all images are degraded by the Gaussian noise $(\mu=0$, $\sigma=2.25)$. In Fig. \ref{NatImg} , segmentation results of ASIC, FCM– AWA, NCM and the proposed method are depicted in each column.
\par We further illustrated the proposed method's performance in image segmentation quantitatively with F-measure \cite{48_pal1991fuzzy} which considers both precision \eqref{Prec} and recall \eqref{Rec} of the segmentation results and is defined by \eqref{Fmesure}: 
\begin{equation}
	F = \dfrac{P.R}{\Psi.P + (1-\Psi).R}
	\label{Fmesure}
\end{equation}
\begin{equation}
P = \dfrac{TP}{TP + FP}
\label{Prec}
\end{equation}
\begin{equation}
R = \dfrac{TP}{TP + FN}
\label{Rec}
\end{equation}
where $\Psi$ is a constant number and is considered as as $0.5$ in \cite{48_pal1991fuzzy}. $P$ is precision, and $R$ is recall rate. $TP$ is the number of correct results, $FP$ is the number of false segmented pixels, and $FN$ is the number of the missed pixels in the result. The $F-measure$ value is in the range of $[0, 1]$, and a larger $F-Measure$ value indicates a higher segmentation accuracy.
\par In Table \ref{NatImgTable} , the $F-measure$ values of each method applied on four natural images are reported. The proposed method achieved the highest F-Measure of 0.93 in $woman$ image and 0.85, 0.88 and 0.85 in $rice$, $eight$ and $lena$ images respectively, which outperforms other methods. Therefore, experimental results show that the proposed method yields more reasonable segmentations than the compared methods quantitatively and qualitatively.
\begin{table}[]
	\centering
	\caption{F-measure values for NCM, FCM–AWA, ASIC and the proposed method.}
	\label{NatImgTable}
	\begin{tabular}{|l|l|l|l|l|}
		\hline
		& ASIC   & FCMAWA & NCM    & Proposed Method \\ \hline
		 rice  & 0.7312 & 0.7802  & 0.8166 & \textbf{0.8506}          \\ \hline
	     eight & 0.7988 & 0.8307  & 0.8627 & \textbf{0.8835}          \\ \hline
		 Lena  & 0.7371 & 0.7882  & 0.8302 & \textbf{0.8565}          \\ \hline
	     woman & 0.7345 & 0.8064  & 0.8739 & \textbf{0.9347} \\  \hline     
	\end{tabular}
\end{table}
For natural images, FCM–AWA  is set with parameters: $m=2$, $\alpha=50$, $\varepsilon=10^{-3}$, $r=1$, $k_0=0.45$ and $k_1=0.65$.
\subsection{Medical image dataset}
\par
\begin{figure}[H]
	\centering
	\includegraphics[width= 1 \textwidth ,height=10cm]{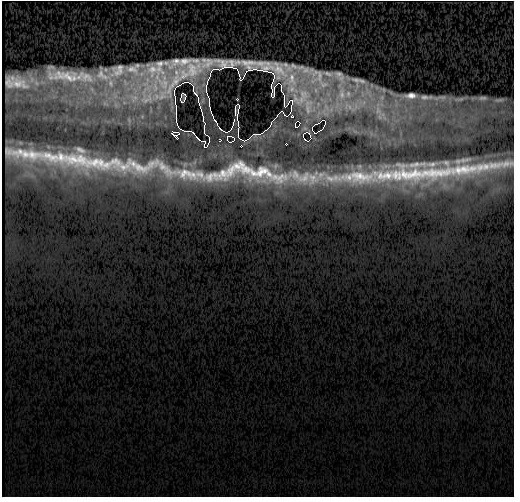}
	\caption{Segmented fluid regions by the proposed clustering method in OCT image with intra-retinal and sub-retinal fluid regions.}
	\vspace{2em}
	\label{SampleForClusteringMed_a}
\end{figure}

\begin{figure}[H]
	\centering
	\includegraphics[width= 1 \textwidth ,height=10cm]{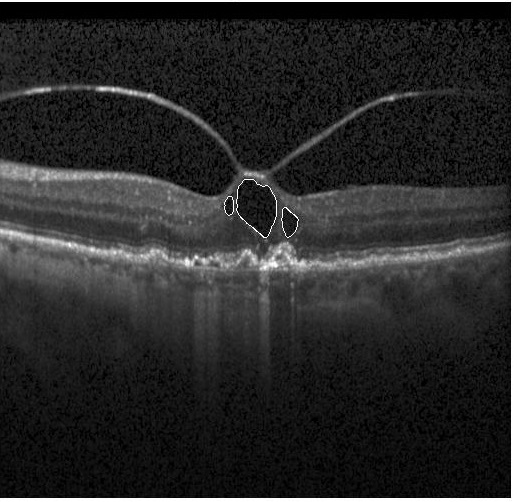}
	\caption{Segmented fluid regions by the proposed clustering method in OCT image with intra-retinal and detached memberance.}
	\vspace{2em}
	\label{SampleForClusteringMed_b}
\end{figure}

\begin{figure}[H]
	\centering
	\includegraphics[width= 1 \textwidth ,height=10cm]{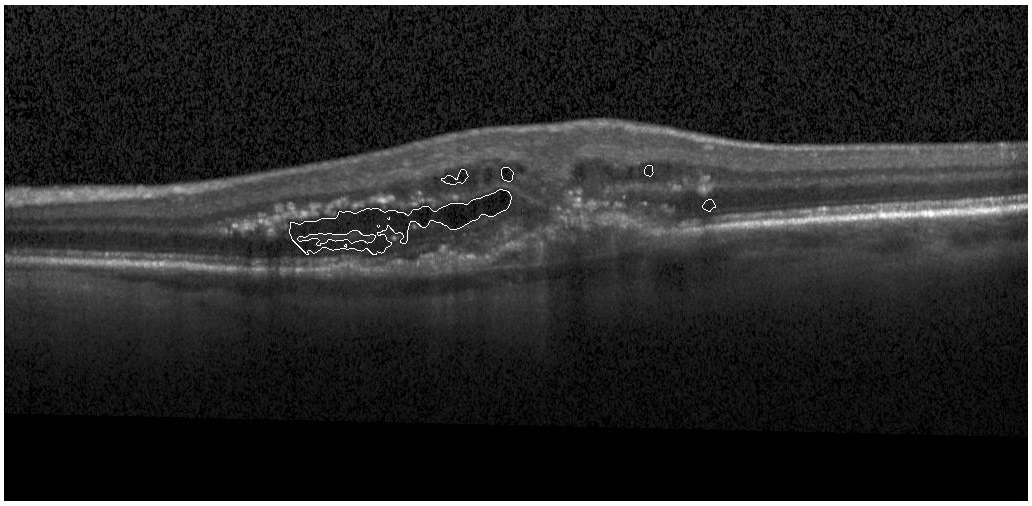}
	\caption{Segmented fluid regions by the proposed clustering method in OCT image with intra-retinal fluid and hard exudate and hyper-reflection regions.}
	\vspace{2em}
	\label{SampleForClusteringMed_c}
\end{figure}

\begin{figure}[H]
	\centering
	\includegraphics[width= 1 \textwidth ,height=10cm]{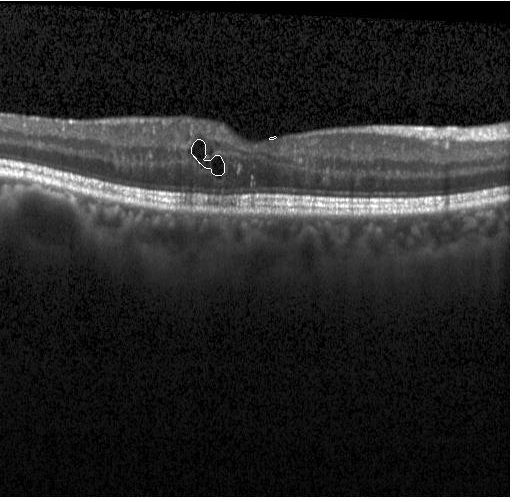}
	\caption{Segmented fluid regions by the proposed clustering method in OCT image with intra-retinal fluid.}
	\vspace{2em}
	\label{SampleForClusteringMed_d}
\end{figure}

 Optical coherence tomography (OCT) is a non-invasive and non-contact imaging method for eye retina which is extensively used clinically for the diagnosis and follow-up of patients with diabetic macular edema (DME) and age-related macular degeneration (AMD). DME and AMD, manifested by fluid regions within the retina and retinal thickening, is caused by fluid leakage from damaged macular blood vessels \cite{36_rashno2017fully,rashno2018octEMBS}.  In this research, the proposed clustering algorithm is applied to cluster OCT images for fluid segmentation. For this purpose, optima OCT dataset including $196$ images from $4$ patients ($49$ images per subject) is used. It should be noted that fluid segmentation in OCT images needs pre-processing, post-processing and layer segmentation. For these steps, we have used the proposed methods in \cite{36_rashno2017fully} since these steps are out of the scope of this research. Table \ref{MedImgTable1} and \ref{MedImgTable2} reports average dice coefficients, precision and sensitivity of the proposed method and methods in \cite{35_rashno2017fully,36_rashno2017fully} and \cite{51_boykov2006graph,52_salah2011multiregion,53_esmaeili2016three,54_de2015machine,55_venhuizena2015vendor}. The proposed method achieved the best average sensitivity of $90.59$ and $89.10$ in comparison with manual expert $1$ and manual expert $2$, respectively.  For dice coefficient and precision measures, methods in \cite{35_rashno2017fully} and \cite{51_boykov2006graph} achieved the best performance. Four samples of OCT images including 1)intra-retinal and sub-retinal fluid, 2)intra-retinal fluid with detached memberance, 3)intera-retinal fluid with hard exudate and  hyper-reflective regions and 4) intra-retinal fluid, are segmented by the proposed clustering method in Figs. \ref{SampleForClusteringMed_a}-\ref{SampleForClusteringMed_d}, respectively.
\begin{table}[H]
	\centering
	\caption{Dice coefficients, sensitivity and precision of different method in comparison with manual expert 1 segmentation results.}
	\label{MedImgTable1}
	\resizebox{\linewidth}{3.5cm}{
	\begin{tabular}{|l|l|l|l|l|l|l|l|l|l|}
		\hline
		    & Sub  & GC\cite{51_boykov2006graph} & KGC\cite{52_salah2011multiregion} & Method in \cite{53_esmaeili2016three} & Method in \cite{54_de2015machine} & Method in \cite{55_venhuizena2015vendor} & Method in \cite{35_rashno2017fully} & Method in \cite{36_rashno2017fully} & Prop. Method \\ \hline
		Dice Coeff.
		& 1    & 73.49      & 80.43       & 71.4               & 61                 & 72                 & 82.96              & \textbf{83.4}               & 81.11           \\ 
		& 2    & 73.9       & 55.1        & 45.49              & 79                 & \textbf{84}                 & 78.11              & 59.5               & 65.8            \\
		& 3    & 78.46      & 75.35       & 69.54              & 43                 & 72                 & \textbf{82.23}              & 71.3               & 77.94           \\
		& 4    & 78.12      & 71.78       & 71.15              & 46                 & 64                 & \textbf{80.75}              & 70.75              & 66.6            \\
		& Ave. & 75.99      & 70.66       & 64.39              & 57.25              & 73                 & \textbf{81.01}              & 70.02              & 72.8            \\ \hline
		Sensitivity & 1    & 70.81      & 82.19       & 72.49              & NA                 & NA                 & 84.43              & 87.6               & \textbf{88.95}           \\
		& 2    & 96.79      & \textbf{99.04}       & 70.45              & NA                 & NA                 & 98.94              & 97.2               & 97.42           \\
		& 3    & 75.72      & 85.13       & 47.38              & NA                 & NA                 & 85.18              & 85.4               & \textbf{85.58}           \\
		& 4    & 78.78      & 80.59       & 77.79              & NA                 & NA                 & 84.49              & 87.2               & \textbf{90.44}           \\
		& Ave. & 80.52      & 86.73       & 67.02              & NA                 & NA                 & 88.26              & 89.35              & \textbf{90.59}           \\ \hline
		Precision   & 1    & \textbf{93}         & 85.06       & 54.87              & NA                 & NA                 & 84.03              & 84.4               & 81.02           \\
		& 2    & 74.36      & 54.18       & 51.12              & NA                 & NA                 & \textbf{78.48}              & 59.5               & 65.52           \\
		& 3    & \textbf{94.89}      & 79.88       & 30.93              & NA                 & NA                 & 85.45              & 77.1               & 82.07           \\
		& 4    & \textbf{96.97}      & 88.62       & 54.98              & NA                 & NA                 & 93.2               & 71.6               & 71.1            \\
		& Ave. & \textbf{89.8}       & 76.93       & 47.97              & NA                 & NA                 & 85.29              & 73.15              & 74.92    \\ \hline      
	\end{tabular}
}
\end{table}
\begin{table}[H]
	\centering
	\caption{Dice coefficients, sensitivity and precision of different method in comparison with manual expert 2 segmentation results..}
	\label{MedImgTable2}
	\resizebox{\linewidth}{3.5cm}{
	\begin{tabular}{|l|l|l|l|l|l|l|l|l|l|}
		\hline
		    & Sub  & GC\cite{51_boykov2006graph} & KGC\cite{52_salah2011multiregion} & Method in \cite{53_esmaeili2016three} & Method in \cite{54_de2015machine} & Method in \cite{55_venhuizena2015vendor} & Method in \cite{35_rashno2017fully} & Method in \cite{36_rashno2017fully} & Prop. Method \\ \hline
		Dice Coeff.  
		& 1    & 72.96      & 79.1        & 68.17              & 56                 & 76                 & \textbf{82.9}               & 81.86              & 80.03           \\
		& 2    & 71.68      & 55.11       & 45.81              & 76                 & \textbf{84}                 & 79.09              & 57.46              & 63.55           \\
		& 3    & \textbf{82.33}      & 79.34       & 65.01              & 42                 & 75                 & 80.36              & 79.09              & 81.09           \\
		& 4    & 77.91      & 71.56       & 72.55              & 45                 & 67                 & \textbf{80.87}              & 65.59              & 70.52           \\
		& Ave. & 76.22      & 71.27       & 62.88              & 54.75              & 75.5               & \textbf{80.8}               & 71                 & 73.79           \\ \hline
		Sensitivity & 1    & 69.95      & 78.56       & 66.75              & NA                 & NA                 & 80.94              & 83.03              & \textbf{84.45}           \\
		& 2    & 92.25      & \textbf{94.54}       & 64.71              & NA                 & NA                 & 94.45              & 92.56              & 92.76           \\
		& 3    & 81.49      & 90.95       & 54.84              & NA                 & NA                 & 90.75              & \textbf{91.57}              & 90              \\
		& 4    & 78.54      & 80.22       & 77.56              & NA                 & NA                 & 83.7               & 86.24              & \textbf{89.21}           \\
		& Ave. & 80.55      & 86.06       & 65.96              & NA                 & NA                 & 87.46              & 88.35              & \textbf{89.1}            \\ \hline
		Precision   & 1    & \textbf{95.71}      & 86.55       & 59.61              & NA                 & NA                 & 87.53              & 87.3               & 83.37           \\
		& 2    & 74.45      & 54.14       & 51.34              & NA                 & NA                 & \textbf{78.58}              & 59.61              & 65.54           \\
		& 3    & \textbf{96.1}       & 79.96       & 37.99              & NA                 & NA                 & 85.48              & 77.21              & 82.79           \\
		& 4    & \textbf{97.24}      & 88.99       & 59.5               & NA                 & NA                 & 93.58              & 72.36              & 71.79           \\
		& Ave. & \textbf{90.87}      & 77.41       & 52.11              & NA                 & NA                 & 86.29              & 74.62              & 75.87     \\ \hline     
	\end{tabular}
}
\end{table}
\section{Discussion }
\par In this section, advantages and disadvantages of the proposed method are discussed. Considering boundary cluster to handle boundary points in methods such as NCM has two side effects which are highly correlated to each other.  First, there are points between boundary cluster and a main cluster such as  points $5$ and $7$ in $X12$; 5, 7, 11 and $13$ in $X19$; 5, 7, 11, 13, 17 and $19$ in $X24$; 3, 6, 9, 25, 28 and $31$ in $X35$. These points are not assigned to a main cluster with a high certainty. The reason is that such points are located in the same distance from the center of the main cluster and the center of boundary cluster. Second, such points displace center of the main clusters. 
\par Consider cluster $1$ in $X35$. Points $23, 5$ and $22$ in $X35$ have the almost same distance from the main cluster center (point $3$) and boundary cluster center (point $6$). Therefore, assigned membership degree of such points to the main cluster and boundary cluster is around $0.50$.  Points $21, 1$ and $20$ have the higher membership degrees$(T_{i,j})$ to the main cluster in comparison with points $23, 5$ and $22$.  In NCM, cluster center is in a direct correlation with $T_{i,j}$:   
\begin{equation}
	c_j\approx\sum_{i=1}^{N}(w_1T_{i,j})^m x_i
\end{equation}
\par Therefore, cluster center around point $3$ is forced to move toward points $21$, $1$ and $20$. This condition is same for other clusters in $X35$. 
\par
\begin{figure}[!t]
	\centering
	\includegraphics[width= 1 \textwidth ,height=7cm]{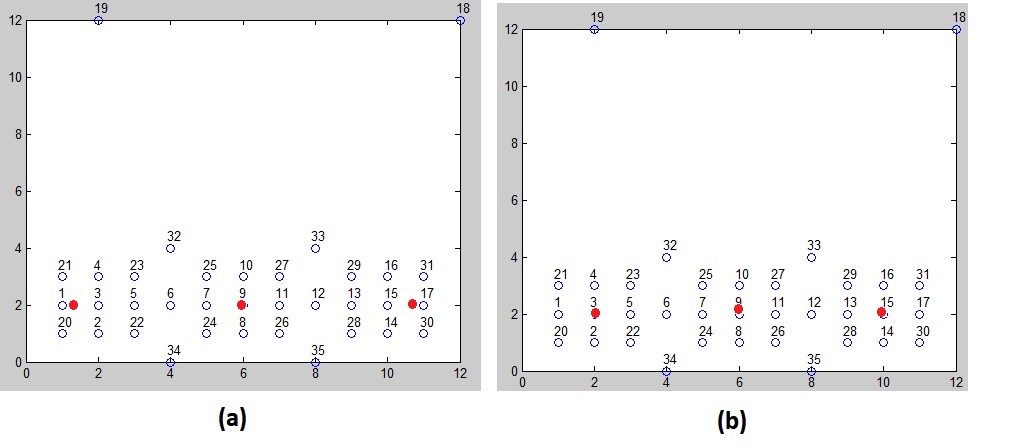}
	\caption{Cluster centers computed by: (a): NCM and (b): the proposed method.}
	\label{ClusCent}
\end{figure}

 When cluster centers are far away from their correct positions (see Fig. \ref{ClusCent}(a)), in the next iteration, since  points $21, 1$ and $20$ are closer to cluster center in comparison with points $23, 5$ and $22$,  their memberships to the main cluster will be stronger. Therefore, these issues are affected by each other in the next iterations of cost function minimization.
\par In the proposed method, these issues are addressed by proposing a cost function and ignoring boundary cluster. As it is depicted in Fig. \ref{ClusCent}, the proposed method is robust and main cluster centers are not forced to be far away from boundary points. Reported results demonstrate how the proposed method handles boundary points and points between boundary points and main cluster centers. 
\par
\begin{figure}[!t]
	\centering
	\includegraphics[width= 0.6 \textwidth ,height=6cm]{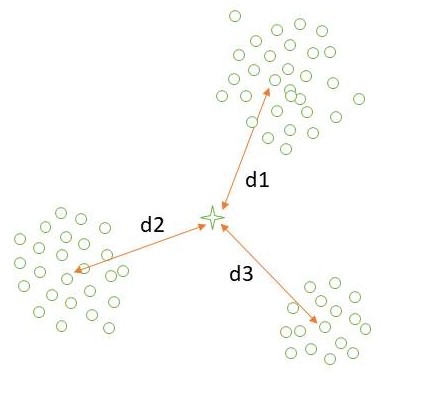}
	\caption{Boundary points.}
	\vspace{2em}
	\label{BoundPoint}
\end{figure}

 As it is shown in Fig. \ref{BoundPoint}, boundary points have a same distance from main cluster centers. In clustering algorithms, when cluster centers are converged; without significant changes in subsequent iterations; for data point $x_i$, term $||x_i - c_j||$ has the same quantity for all cluster centers $(j={1,2,...,K})$. It leads membership degree $x_i$ to cluster center $j$ ($T_{i,j}$) to be almost same for all clusters. We have used this property to distinguish boundary points.
\par The main problem in clustering is determining the number of main clusters (K). This issue is domain-specific and should be determined under expert supervision. Here, in each experiment $K$ is determined from a context knowledge. It should be noted that inappropriate $K$ affects clustering results significantly. Also, although benefits of NS theory and some aspects of data clustering such as indeterminacy is considered in the proposed cost function, cost function minimization suffers from local optimum points. Finally, although the proposed indeterminacy definition in neutrosophic domain is appropriate for density-based and center-based clustering, it is not working well on non-density cases such as contiguous-based and nonlinear-shape clusters.  
\section{Conclusion}
\par In this research, an effective clustering method was proposed in NS domain. For this task, data indeterminacy was proposed based on density properties of data in NS domain to control outlier and boundary points followed by proposing a cost function in NS domain. Two types of clusters including main clusters and noisy cluster are considered in the proposed cost function.  Experiments on different datasets including diamond datasets; UCI datasets, artificial and natural images; and medical images showed that the proposed method not only handle outlier and boundary points but also  outperforms existing methods in both scatter data clustering and image segmentation. Future efforts will be directed towards introducing indeterminacy in NS domain to supervised methods such as deep convolutional neural networks. Future efforts will be also directed towards proposing methods in neutrosophic domain for handling contiguous-based and nonlinear-shape clusters.

\section*{ACKNOWLEDGMENT}
Authors would like to thank  Dr. Abdolreza Rashno for recommending us to evaluate our method on retina OCT images and providing source code for pre-processing and post-processing steps in OCT fluid segmentation.
\section*{References}

\bibliography{manuscipt}

\begin{thebibliography}{10}
\expandafter\ifx\csname url\endcsname\relax
  \def\url#1{\texttt{#1}}\fi
\expandafter\ifx\csname urlprefix\endcsname\relax\def\urlprefix{URL }\fi
\expandafter\ifx\csname href\endcsname\relax
  \def\href#1#2{#2} \def\path#1{#1}\fi

\bibitem{ClusteringSurveyxu2005survey}
R.~Xu, D.~Wunsch, Survey of clustering algorithms, IEEE Transactions on neural
  networks 16~(3) (2005) 645--678.

\bibitem{ESWA1_khan2013cluster}
S.~S. Khan, A.~Ahmad, Cluster center initialization algorithm for k-modes
  clustering, Expert Systems with Applications 40~(18) (2013) 7444--7456.

\bibitem{ESW2Aheloulou2017automatic}
I.~Heloulou, M.~S. Radjef, M.~T. Kechadi, Automatic multi-objective clustering
  based on game theory, Expert Systems with Applications 67 (2017) 32--48.

\bibitem{ESWA3saha2016brain}
S.~Saha, A.~K. Alok, A.~Ekbal, Brain image segmentation using semi-supervised
  clustering, Expert Systems with Applications 52 (2016) 50--63.

\bibitem{ESWA4ramon2017cluster}
R.~Ramon-Gonen, R.~Gelbard, Cluster evolution analysis: Identification and
  detection of similar clusters and migration patterns, Expert Systems with
  Applications 83 (2017) 363--378.

\bibitem{6-7_baraldi1999survey}
A.~Baraldi, P.~Blonda, A survey of fuzzy clustering algorithms for pattern
  recognition. i, IEEE Transactions on Systems, Man, and Cybernetics, Part B
  (Cybernetics) 29~(6) (1999) 778--785.

\bibitem{19_guo2015ncm}
Y.~Guo, A.~Sengur, Ncm: Neutrosophic c-means clustering algorithm, Pattern
  Recognition 48~(8) (2015) 2710--2724.

\bibitem{23_webb2003statistical}
A.~R. Webb, Statistical pattern recognition, John Wiley \& Sons, 2003.

\bibitem{24_wagstaff2001constrained}
K.~Wagstaff, C.~Cardie, S.~Rogers, S.~Schr{\"o}dl, et~al., Constrained k-means
  clustering with background knowledge, in: ICML, Vol.~1, 2001, pp. 577--584.

\bibitem{25_arthur2007k}
D.~Arthur, S.~Vassilvitskii, k-means++: The advantages of careful seeding, in:
  Proceedings of the eighteenth annual ACM-SIAM symposium on Discrete
  algorithms, Society for Industrial and Applied Mathematics, 2007, pp.
  1027--1035.

\bibitem{27_arora1998approximation}
S.~Arora, P.~Raghavan, S.~Rao, Approximation schemes for euclidean k-medians
  and related problems, in: Proceedings of the thirtieth annual ACM symposium
  on Theory of computing, ACM, 1998, pp. 106--113.

\bibitem{4_zhang2010neutrosophic}
M.~Zhang, L.~Zhang, H.~Cheng, A neutrosophic approach to image segmentation
  based on watershed method, Signal Processing 90~(5) (2010) 1510--1517.

\bibitem{5_bezdek1981objective}
J.~C. Bezdek, Objective function clustering, in: Pattern recognition with fuzzy
  objective function algorithms, Springer, 1981, pp. 43--93.

\bibitem{8_menard2000fuzzy}
M.~M{\'e}nard, C.~Demko, P.~Loonis, The fuzzy c+ 2-means: solving the ambiguity
  rejection in clustering, Pattern recognition 33~(7) (2000) 1219--1237.

\bibitem{9_yang2008alpha}
M.-S. Yang, K.-L. Wu, J.-N. Hsieh, J.~Yu, Alpha-cut implemented fuzzy
  clustering algorithms and switching regressions, IEEE Transactions on
  Systems, Man, and Cybernetics, Part B (Cybernetics) 38~(3) (2008) 588--603.

\bibitem{10_yu2004analysis}
J.~Yu, Q.~Cheng, H.~Huang, Analysis of the weighting exponent in the fcm, IEEE
  Transactions on Systems, Man, and Cybernetics, Part B (Cybernetics) 34~(1)
  (2004) 634--639.

\bibitem{11_gustafson1979fuzzy}
D.~E. Gustafson, W.~C. Kessel, Fuzzy clustering with a fuzzy covariance matrix,
  in: Decision and Control including the 17th Symposium on Adaptive Processes,
  1978 IEEE Conference on, IEEE, 1979, pp. 761--766.

\bibitem{12_krishnapuram1993possibilistic}
R.~Krishnapuram, J.~M. Keller, A possibilistic approach to clustering, IEEE
  transactions on fuzzy systems 1~(2) (1993) 98--110.

\bibitem{13_pal1997mixed}
N.~R. Pal, K.~Pal, J.~C. Bezdek, A mixed c-means clustering model, in: Fuzzy
  Systems, 1997., Proceedings of the Sixth IEEE International Conference on,
  Vol.~1, IEEE, 1997, pp. 11--21.

\bibitem{14_roubens1978pattern}
M.~Roubens, Pattern classification problems and fuzzy sets, Fuzzy sets and
  systems 1~(4) (1978) 239--253.

\bibitem{15_hathaway1989relational}
R.~J. Hathaway, J.~W. Davenport, J.~C. Bezdek, Relational duals of the c-means
  clustering algorithms, Pattern recognition 22~(2) (1989) 205--212.

\bibitem{16_sen1998clustering}
S.~Sen, R.~Dave, Clustering of relational data containing noise and outliers,
  in: Fuzzy Systems Proceedings, 1998. IEEE World Congress on Computational
  Intelligence., The 1998 IEEE International Conference on, Vol.~2, IEEE, 1998,
  pp. 1411--1416.

\bibitem{17_masson2008ecm}
M.-H. Masson, T.~Denoeux, Ecm: An evidential version of the fuzzy c-means
  algorithm, Pattern Recognition 41~(4) (2008) 1384--1397.

\bibitem{18_masson2009recm}
M.-H. Masson, T.~Den{\oe}ux, Recm: Relational evidential c-means algorithm,
  Pattern Recognition Letters 30~(11) (2009) 1015--1026.

\bibitem{Clust1li2018clustering}
X.~Li, Q.~Han, B.~Qiu, A clustering algorithm using skewness-based boundary
  detection, Neurocomputing 275 (2018) 618--626.

\bibitem{Clust2cui2018subspace}
G.~Cui, X.~Li, Y.~Dong, Subspace clustering guided convex nonnegative matrix
  factorization, Neurocomputing 292 (2018) 38--48.

\bibitem{Clust3tong2018efficient}
Q.~Tong, X.~Li, B.~Yuan, Efficient distributed clustering using boundary
  information, Neurocomputing 275 (2018) 2355--2366.

\bibitem{Clust4hammou2018convexity}
F.~Hammou, K.~Hammouche, J.-G. Postaire, Convexity dependent anisotropic
  diffusion for mode detection in cluster analysis, Neurocomputing.

\bibitem{Clust5saxena2017review}
A.~Saxena, M.~Prasad, A.~Gupta, N.~Bharill, O.~P. Patel, A.~Tiwari, M.~J. Er,
  W.~Ding, C.-T. Lin, A review of clustering techniques and developments,
  Neurocomputing 267 (2017) 664--681.

\bibitem{26_smarandache1995neutrosophic}
F.~Smarandache, Neutrosophic logic and set.

\bibitem{1_smarandache2003unifying}
F.~Smarandache, A Unifying Field in Logics: Neutrosophic Logic. Neutrosophy,
  Neutrosophic Set, Neutrosophic Probability: Neutrosophic Logic: Neutrosophy,
  Neutrosophic Set, Neutrosophic Probability, Infinite Study, 2003.

\bibitem{2-29_guo2009new}
Y.~Guo, H.-D. Cheng, New neutrosophic approach to image segmentation, Pattern
  Recognition 42~(5) (2009) 587--595.

\bibitem{3_zhang2010neutrosophic}
M.~Zhang, L.~Zhang, H.~Cheng, A neutrosophic approach to image segmentation
  based on watershed method, Signal Processing 90~(5) (2010) 1510--1517.

\bibitem{30_sengur2011color}
A.~Sengur, Y.~Guo, Color texture image segmentation based on neutrosophic set
  and wavelet transformation, Computer Vision and Image Understanding 115~(8)
  (2011) 1134--1144.

\bibitem{31_heshmati2016scheme}
A.~Heshmati, M.~Gholami, A.~Rashno, Scheme for unsupervised colour--texture
  image segmentation using neutrosophic set and non-subsampled contourlet
  transform, IET Image Processing 10~(6) (2016) 464--473.

\bibitem{32_guo2017efficient}
Y.~Guo, Y.~Akbulut, A.~{\c{S}}eng{\"u}r, R.~Xia, F.~Smarandache, An efficient
  image segmentation algorithm using neutrosophic graph cut, Symmetry 9~(9)
  (2017) 185.

\bibitem{salafian2018automatic}
B.~Salafian, R.~Kafieh, A.~Rashno, M.~Pourazizi, S.~Sadri, Automatic
  segmentation of choroid layer in edi oct images using graph theory in
  neutrosophic space, arXiv preprint arXiv:1812.01989.

\bibitem{33_guo2014novel}
Y.~Guo, A.~{\c{S}}eng{\"u}r, J.~Ye, A novel image thresholding algorithm based
  on neutrosophic similarity score, Measurement 58 (2014) 175--186.

\bibitem{34_guo2014novel}
Y.~Guo, A.~{\c{S}}eng{\"u}r, A novel image edge detection algorithm based on
  neutrosophic set, Computers \& Electrical Engineering 40~(8) (2014) 3--25.

\bibitem{rashno2017refined}
A.~Rashno, F.~Smarandache, S.~Sadri, Refined neutrosophic sets in content-based
  image retrieval application, in: Machine Vision and Image Processing (MVIP),
  2017 10th Iranian Conference on, IEEE, 2017, pp. 197--202.

\bibitem{46_rashno2017content}
A.~Rashno, S.~Sadri, Content-based image retrieval with color and texture
  features in neutrosophic domain, in: Pattern Recognition and Image Analysis
  (IPRIA), 2017 3rd International Conference on, IEEE, 2017, pp. 50--55.

\bibitem{35_rashno2017fully}
A.~Rashno, B.~Nazari, D.~D. Koozekanani, P.~M. Drayna, S.~Sadri, H.~Rabbani,
  K.~K. Parhi, Fully-automated segmentation of fluid regions in exudative
  age-related macular degeneration subjects: Kernel graph cut in neutrosophic
  domain, PloS one 12~(10) (2017) e0186949.

\bibitem{36_rashno2017fully}
A.~Rashno, D.~D. Koozekanani, P.~M. Drayna, B.~Nazari, S.~Sadri, H.~Rabbani,
  K.~K. Parhi, Fully-automated segmentation of fluid/cyst regions in optical
  coherence tomography images with diabetic macular edema using neutrosophic
  sets and graph algorithms, IEEE Transactions on Biomedical Engineering.

\bibitem{37_rashno2017automated}
A.~Rashno, K.~K. Parhi, B.~Nazari, S.~Sadri, H.~Rabbani, P.~Drayna, D.~D.
  Koozekanani, Automated intra-retinal, sub-retinal and sub-rpe cyst regions
  segmentation in age-related macular degeneration (amd) subjects,
  Investigative Ophthalmology \& Visual Science 58~(8) (2017) 397--397.

\bibitem{38_parhi2017automated}
K.~K. Parhi, A.~Rashno, B.~Nazari, S.~Sadri, H.~Rabbani, P.~Drayna, D.~D.
  Koozekanani, Automated fluid/cyst segmentation: A quantitative assessment of
  diabetic macular edema, Investigative Ophthalmology \& Visual Science 58~(8)
  (2017) 4633--4633.

\bibitem{kohler2017correlation}
J.~Kohler, A.~Rashno, K.~K. Parhi, P.~Drayna, S.~Radwan, D.~D. Koozekanani,
  Correlation between initial vision and vision improvement with automatically
  calculated retinal cyst volume in treated dme after resolution, Investigative
  Ophthalmology \& Visual Science 58~(8) (2017) 953--953.

\bibitem{39_guo2017novel}
Y.~Guo, A.~S. Ashour, B.~Sun, A novel glomerular basement membrane segmentation
  using neutrsophic set and shearlet transform on microscopic images, Health
  information science and systems 5~(1) (2017) 15.

\bibitem{40_guo2017retinal}
Y.~Guo, {\"U}.~Budak, A.~{\c{S}}eng{\"u}r, F.~Smarandache, A retinal vessel
  detection approach based on shearlet transform and indeterminacy filtering on
  fundus images, Symmetry 9~(10) (2017) 235.

\bibitem{41_siri2017combined}
S.~K. Siri, M.~V. Latte, Combined endeavor of neutrosophic set and chan-vese
  model to extract accurate liver image from ct scan, Computer methods and
  programs in biomedicine 151 (2017) 101--109.

\bibitem{42_sirinovel}
S.~K. Siri, M.~V. Latte, A novel approach to extract exact liver image boundary
  from abdominal ct scan using neutrosophic set and fast marching method,
  Journal of Intelligent Systems.

\bibitem{43_lotfollahi2017segmentation}
M.~Lotfollahi, M.~Gity, J.~Y. Ye, A.~M. Far, Segmentation of breast ultrasound
  images based on active contours using neutrosophic theory, Journal of Medical
  Ultrasonics (2017) 1--8.

\bibitem{44_akbulut2017ns}
Y.~Akbulut, A.~Sengur, Y.~Guo, F.~Smarandache, Ns-k-nn: Neutrosophic set-based
  k-nearest neighbors classifier, Symmetry 9~(9) (2017) 179.

\bibitem{45_dhar2017accurate}
S.~Dhar, M.~K. Kundu, Accurate segmentation of complex document image using
  digital shearlet transform with neutrosophic set as uncertainty handling
  tool, Applied Soft Computing 61 (2017) 412--426.

\bibitem{20_akbulut2017kncm}
Y.~Akbulut, A.~{\c{S}}eng{\"u}r, Y.~Guo, K.~Polat, Kncm: Kernel neutrosophic
  c-means clustering, Applied Soft Computing 52 (2017) 714--724.

\bibitem{FCM_bezdek1984fcm}
J.~C. Bezdek, R.~Ehrlich, W.~Full, Fcm: The fuzzy c-means clustering algorithm,
  Computers \& Geosciences 10~(2-3) (1984) 191--203.

\bibitem{48_pal1991fuzzy}
S.~K. Pal, Fuzzy tools for the management of uncertainty in pattern
  recognition, image analysis, vision and expert systems, International journal
  of systems science 22~(3) (1991) 511--549.

\bibitem{49_kang2009novel}
J.~Kang, L.~Min, Q.~Luan, X.~Li, J.~Liu, Novel modified fuzzy c-means algorithm
  with applications, Digital signal processing 19~(2) (2009) 309--319.

\bibitem{51_boykov2006graph}
Y.~Boykov, G.~Funka-Lea, Graph cuts and efficient nd image segmentation,
  International journal of computer vision 70~(2) (2006) 109--131.

\bibitem{52_salah2011multiregion}
M.~B. Salah, A.~Mitiche, I.~B. Ayed, Multiregion image segmentation by
  parametric kernel graph cuts, IEEE Transactions on Image Processing 20~(2)
  (2011) 545--557.

\bibitem{53_esmaeili2016three}
M.~Esmaeili, A.~M. Dehnavi, H.~Rabbani, F.~Hajizadeh, Three-dimensional
  segmentation of retinal cysts from spectral-domain optical coherence
  tomography images by the use of three-dimensional curvelet based k-svd,
  Journal of medical signals and sensors 6~(3) (2016) 166.

\bibitem{54_de2015machine}
L.~de~Sisternes, J.~Hong, T.~Leng, D.~L. Rubin, A machine learning aproach for
  device-independent automated segmentation of retinal cysts in spectral domain
  optical cohorence tomography images.

\bibitem{55_venhuizena2015vendor}
F.~G. Venhuizena, M.~Grinsvena, C.~B. Hoyngb, T.~Theelenb, B.~Ginnekena, C.~I.
  Sancheza, Vendor independent cyst segmentation in retinal sd-oct volumes
  using a combination of multiple scale convolutional neural networks, Medical
  Image Computing and Computer Assisted Intervention-Challenge on Retinal Cyst
  Segmentation.

\bibitem{FCM_new}
M.~M{\'e}nard, C.~Demko, P.~Loonis, The fuzzy c+ 2-means: solving the ambiguity
  rejection in clustering, Pattern recognition 33~(7) (2000) 1219--1237.

\bibitem{PCM_new}
M.~Roubens, Pattern classification problems and fuzzy sets, Fuzzy sets and
  systems 1~(4) (1978) 239--253.

\bibitem{PFCM_new}
N.~R. Pal, K.~Pal, J.~M. Keller, J.~C. Bezdek, A possibilistic fuzzy c-means
  clustering algorithm, IEEE transactions on fuzzy systems 13~(4) (2005)
  517--530.

\bibitem{HPFCM_new}
N.-E. El~Harchaoui, M.~A. Kerroum, A.~Hammouch, M.~Ouadou, D.~Aboutajdine,
  Unsupervised approach data analysis based on fuzzy possibilistic clustering:
  application to medical image mri, Computational intelligence and neuroscience
  2013 (2013) 10.

\bibitem{rashno2018octEMBS}
A.~Rashno, D.~D. Koozekanani, K.~K. Parhi, Oct fluid segmentation using graph
  shortest path and convolutional neural network, in: 2018 40th Annual
  International Conference of the IEEE Engineering in Medicine and Biology
  Society (EMBC), IEEE, 2018, pp. 3426--3429.

\end{thebibliography}

\end{document}